\documentstyle[prb,aps,epsf]{revtex}

\begin{document}
\draft

\twocolumn[\hsize\textwidth\columnwidth\hsize\csname @twocolumnfalse\endcsname

\title
{
An analysis of
photoemission and inverse photoemission spectra of Si(111) and
sulphur-passivated InP(001) surfaces
}

\author
{
 M.W.C. Dharma-wardana,$^{1,}$\cite{byline1} Z. Tian,$^{1,2}$ Z.H. Lu,$^1$
and L.J. Lewis$^2$
}

\address
{
$^1$Institute for Microstructural Sciences, National Research Council,
Ottawa, Canada K1A 0R6 \\
$^2$D\'epartement de physique et Groupe de recherche en physique et technologie
des couches minces (GCM), Universit\'e de Montr\'eal, Case Postale 6128,
Succursale Centre-Ville, Montr\'eal, Qu\'ebec, Canada H3C 3J7 \\
}

\date{\today}

\maketitle


\begin{abstract}

Photoemission (PES) and inverse-photoemission spectra (IPES)
for the sulphur-passivated
InP(001) surface are compared with theoretical predictions based on
density-functional calculations. As a test case for our methods, we
also present a corresponding study of the better known Si(111) surface.
 The reported spectra for InP(001)-S agree well with the calculated ones
 if the surface is
assumed to consist of a mixture of two phases, namely, the fully
S-covered $(2\times2)$-reconstructed structure, which contains four S atoms in
the surface unit-cell, and a $(2\times2)$ structure containing two S and two
P atoms per unit cell. The latter has recently been identified in
total-energy calculations as well as in core-level spectra of
 S-passivated InP(001) surfaces under annealing.
The experimental IPES for Si(111)-$(2\times1)$ is in excellent agreement
with the calculations. The comparison of the experimental-PES with
our calculations provides additional considerations regarding the
nature of the sample surface. It is also found that the
commonly-used density-of-states approximation to the photo- and inverse-
photoemission spectra is not valid for these systems.

\end{abstract}

\vskip2pc]

\narrowtext

\section{INTRODUCTION}
\label{intro}
The sulphur-treated indium phospide surface, InP(001)-S, has recently
attracted much attention, where the effort towards a microscopic
understanding of the physics has been spurred by its
technological importance in
 surface passivation problems.\cite{sandroff89,Tao92,Science 93}
 This surface has been studied
experimentally using a variety of methods, including X-ray absorption
near-edge spectroscopy,\cite{graham} Raman spectroscopy,\cite{jin}
core-level spectroscopy (CLS),\cite{fukuda,chasse,tian} photoemission
(PES) and inverse photoemission (IPES) spectroscopy,\cite{mclean} low-energy
electron diffraction (LEED),\cite{warren} and scanning tunneling microscopy
(STM).\cite{norton} Theoretically, it has been examined in detail using
total energy minimization based on
 density-functional theory (DFT).\cite{jin,tian}

Early studies suggested that the InP(001)-S surface is a $(1\times1)$
structure that follows the bulk (zinc-blende) pattern, terminated with the
sulphur atoms on the bridge sites (see for instance Refs.\
\onlinecite{graham} and \onlinecite{warren}). The actual situation, however,
has turned out to be much more complicated. Thus, LEED data on samples
annealed at 350$^{\circ}$C appeared to have a clear $(2\times1)$
pattern.\cite{mclean} However, Raman experiments and theoretical
calculations indicated\cite{jin} that the ``as-prepared'' low-temperature
material arranges into a novel $(2\times2)$ structure with {\em two} types of
sulphur on the surface, as displayed in Fig.\ 1. This structure has four
sulphur atoms on the surface unit-cell, belonging to two distinct sublayers
about 0.2 \AA\ apart; the bottom sublayer is dimerized, i.e., contains a
S$_2$ ``molecule'', while the top sublayer consists of two S atoms in a
monomer state. We will refer to this surface as the S$_2$SS structure.

Further theoretical calculations, combined with CLS measured as a
function of the annealing temperature of the surface, showed
that other surface structures were allowed within a given
thermal window. The most important of these
forms {\it via} the exchange of surface-sulphur
atoms with bulk-phosphorous atoms which migrate
to the surface.\cite{tian} It was
demonstrated that the most-likely structure in the 300--400$^{\circ}$C
annealing range is a $(2\times2)$-reconstruction containing essentially
 {\em one} type of
sulphur, with the surface layer containing equal amounts of S and P atoms
that form tight S--P bonds (Fig. 2) which are slightly tilted
 so that the P-atoms are on top of the surface.
 We will refer to this structure as
(SP)$_2$. At much higher temperatures, all the surface sulphur gets replaced
by phosphorous, leading to a P-terminated sulphur-free
InP(001) surface. Although the
(SP)$_2$ structure is a $(2\times2)$-reconstruction, it could easily appear
to be $(2\times1)$ if LEED fails to distinguish between S and P atoms. Thus,
we are lead to a picture wherein the InP(001)-S surface is actually a system
which could contain a mixture of S$_2$SS, (SP)$_2$, and the P-terminated
InP(001) structure, depending on the kinetics imposed by the annealing
conditions.

The objective of this paper is to examine the reported PES and IPES data for
the samples annealed at 350$^{\circ}$C by comparing them with
first-principles
DFT calculations of PES and IPES for the energetically-likely
surface structures. It turns out that the best fit to the data is
obtained by assuming that the surface is in a mixed phase  consisting
of S$_2$SS and (SP)$_2$ patches. In order to ``benchmark''
our calculations, we present similar calculations for the
$(2\times1)$-reconstructed Si(111) surface. The spectra for
Si(111)-$(2\times1)$ available in the literature are for samples prepared by
cleavage in {\it vacuuo}.\cite{himpsel}
These spectra also contain various features which could arise
from surface imperfections or even the presence of
some admixture of the more stable $(7\times7)$
reconstruction. We have examined these issues
within our analysis of the reported PES-spectra
of Si(111),
and they serve as a testing ground of the theoretical
methods used here.

The plan of the paper is as follows. We first present, in Section
\ref{theory}, the essentials of the theory. In Section \ref{results}, we
start by discussing our results for the Si(111) surface and compare them with
the experimental data of Himpsel {\em et al.}\cite{himpsel} In particular, we
emphasize that that the PES and IPES spectra are {\em not} well accounted for
by the density of states (DOS) of occupied or unoccupied levels. However,
good agreement with theory is obtained when the calculation is done using the
transition matrix elements obtained from the Kohn-Sham energy bands.
 We then proceed to a study of the InP(001)-S case
and find that it is not possible to explain the experimental PES
data in terms of a single-phase surface; however, we find good
agreement if the surface is assumed to consist of a mixture of S$_2$SS and
(SP)$_2$ phases which have been identified in annealing experiments.

\section{THEORY}
\label{theory}

The geometrical details of the S$_2$SS and (SP)$_2$ structures described
above were determined within the framework of DFT
total-energy minimization calculations; full details can be found in Refs.\
\onlinecite{jin} and \onlinecite{tian}. Given these structures, as in our
previous studies, we construct supercells containing typically 16 atomic
layers and the equivalent of 5 layers of vacuum. In the case of the
Si(111) slab, the top and bottom 4 layers of the slab have the experimental
geometry of the 2$\times$1 reconstruction, while the middle eight layers
have the bulk-Si geometry. These supercells are then used to calculate the
electronic energy states (band structure) of these systems. We use, for this
purpose, the all-electron, full-potential-linear-muffin-tin-orbital (FP-LMTO)
method;\cite{methfessel} in this formalism, the wavefunctions $\psi(\vec r)$
are of the form:
   \begin{equation}
   \label{psi}
   \psi_i^{\vec k}(\vec r)=\sum_{t,L,K}C^n_{t,l,K}\chi^{\vec k}_{t,L}
   (E,K,{\vec r}),
   \end{equation}
where $n$ is the band index and ${\vec k}$ is the electron wavevector; $t$,
$L$, $K$, and $E$ are, respectively, the index of the atomic species in the
unit cell, the angular momentum, the Hankel function expansion parameter
(with $K^2$=$-|\epsilon|$), and the reference energy $E$ of the Bloch basis
$\chi^{\vec k}_{t,L}$.\cite{methfessel}

The transition probability between initial and final states, $\psi_i(\vec r)$
and $\psi_f(\vec r)$, induced by the electron-photon interaction $H_{int}$ is
given by the Fermi golden rule as
   \begin {equation}
   \label{fgr}
   W_{if}=\frac{2\pi}{\hbar} |\langle f\mid H_{int}\mid i\rangle|^2
   \delta(E_f-E_i-\hbar \omega)
   \end{equation}
where $\hbar \omega$ is the photon energy. For PES, electrons with final
energy $E_f$ will escape from the sample with kinetic energy $E_{kin}$ =
$E_f-E_{F}-\phi$, where $\phi$ is the workfunction. The interaction
Hamiltonian  is ${\vec A}\cdot{\vec p}$, where ${\vec A}$
is the vector potential of the photon field. In many cases, it is necessary
to replace the external field due to the vector potential ${\vec A}$
of the photons by
 the total field inclusive of
induced fields using a Sternheimer-type procedure, as shown for instance by
Zangwill and Soven.\cite{zangwill} However, in our calculations, we restrict
ourselves to the spectrum given by the simplest form of the Fermi golden
rule. The velocity form of the matrix was used in the numerical
calculations. The total photoelectron current per unit solid angle
in PES is then
   \begin{equation}
   J^{PES}(E_{kin},\omega) \propto \surd E_{kin}\sum_{i}W_{if}.
   \end{equation}
The expression for the transition probability for IPES is identical to PES;
the IPES current is given by
   \begin{equation}
   J^{IPES}(E_{kin},\omega) \propto [1/\surd E_{kin}]\sum_{f}W_{fi}\\
   \end{equation}
where, in this case, $E_{kin} = E_i-E_F-\phi$.

Rather than evaluating explicitly the transition matrix elements $\langle
f\mid {\vec A}\cdot{\vec p}\mid i\rangle$, it is often assumed, for
simplicity, that these are constant in the range of the energies concerned.
In this approximation, the PES/IPES intensities become proportional to the
DOS of occupied/unoccupied levels. Another approximation that is made in this
context is to assume that the final electron state
 in PES (or the initial electron
state in IPES) is a single plane-wave state. In reality, the final state
in PES is probably more like
a superposition of plane-waves dictated by the crystal potential
and is best represented by the full, self-consistent, Kohn-Sham eigenfunction
corresponding to the energy of the final state. It should be noted that the
Kohn-Sham eigenvalues are well known to be a poor approximation to the
quasi-particle energies, but the eigenfunctions themselves are generally a
good approximation to the quasi-particle wave-functions. Thus, in the present
study, we treat the PES and IPES calculations as separate problems dealing
with separate experiments (as they indeed are), and do not attempt to relate
them to each other, or to the experimental surface bandgaps; rather, we
determine explicitely the Kohn-Sham eigenstates appropriate to the
occupied/unoccupied levels, and use these to evaluate in detail the
transition matrix elements of Eqs.\ (3) and (4). In fact, the
DOS profile is found to yield  poor approximations to the measured spectra.
This fact is of some importance since experimental PES/IPES results
are often presented as a ``mapping''
of the ``bandstructure'' of a material, and theoretical bandstructures are
often ``corrected'' to fit in with the PES-dispersions, without
deconvoluting the effect of the matrix elements and the non-plane wave
character of the final (or initial ) states in PES (or IPES).

In our numerical calculations, the $\delta$-functions appearing in Eq.\
\ref{fgr} and related equations will be replaced by a Lorentzian functions of
width 0.2 eV. In calculating the matrix elements, the wavefunction given in
Eq.\ \ref{psi} was expanded in plane-waves for ease of computation. An energy
cutoff of 20 Ry was found to be sufficient for the spectral range studied,
while test calculations using 30 Ry were performed in order to establish
convergence in the expansions. Calculations were carried out for bulk Si,
Si(111)-(2$\times$1), the S$_2$SS and (SP)$_2$ surface phases, and bulk InP.
Since we are dealing with admixtures of several phases, i.e., surfaces which
may only have short-range order, calculations are reported only for the
${\bar\Gamma}$ point of the surface Brillouin zone. It was verified that
increasing the number of ``bulk-like'' layers in the simulation supercells
helped to augment the bulk-like signal in the theoretical spectra, arising
from the projected bulk DOS at the surface's ${\bar\Gamma}$ point. Indeed,
since light penetrates many atomic layers, the experimental spectra contain,
to some extent, contributions from the bulk, as well as from resonances due
to interactions among the surface and bulk excitations.
In fact, one effect of coupling a surface slab
to the bulk is to broaden all the peaks. This is similar to the effect of the
Lorentzian broadening of the delta-functions used here. Also,
by considering a
sufficient number of bulk-like layers, we ensure that such effects are
represented in the theory.
Thus we  checked for the convergence of
the density of states in each layer as a function of slab thickness by
considering systems containing up to 16 layers. The surface unit cell
was always taken to be (2$\times$2); that is, no calculations were
performed for the P-terminated InP surface free of sulphur resulting from
high temperature annealing, which has a (2$\times$4) structure, or for the
Si(111)-(7$\times$7) surface which is also considered in some of the
 discussions.

\section{RESULTS AND DISCUSSION}
\label{results}

In order to establish the reliability  of our approach, we first consider
the well-known Si(111) surface as a test case.
We compare the results available in
the literature with our theoretical calculations and return to the problem
of interest here, viz.\ the InP(001)-S surface.

\subsection{PES and IPES for the Si(111) surface}
\label{si-pes}

The Si(111)-(2$\times$1) surface was extensively studied in the early
eighties and its spectroscopy played a pivotal role in eliminating the early
models in favour of Pandey's $\pi$-bonded chain
structure.\cite{himpsel,pandy,northrup,hansson,uhrberg}
 We use here some of the
published spectra for the nominal Si(111)-(2$\times$1) surface as a benchmark
for our calculations.

Fig.\ \ref{sicbfig} compares our calculations for the IPES using the
Kohn-Sham transition matrix elements (solid line), and DOS for unoccupied
levels (dashed line), with the experimental results of Himpsel\cite{himprep}
(squares) for the Si(111)-(2$\times$1) surface. Here we used a computational
supercell consisting of a 16-layer slab of Si and 5 layers of vacuum. The
absolute position of the theoretical curve was displaced to conform with the
energy scale reported in the experiment. The Fermi energy of the (2$\times$1)
 surface is given, experimentally, to be
0.33 eV above the valence band maximum and this has been used in positioning
the experimental spectrum. It is clear that the IPES
calculation using the simple Fermi golden rule and the {\em full} matrix
elements is in excellent agreement with experiment, while the corresponding
DOS calculation is a poor representation of the experimental spectrum.

In Fig.\ \ref{sivbx},  we present our calculated
PES (dashed line) and DOS (dotted
line) results for the Si(111)-(2$\times$1) surface, the calculated bulk
PES signal (dot-dashed lined), as well as the
experimental results of Himpsel {\em et al.}\cite{himpsel}
We have not subtracted off the secondary-electron
background in displaying the
experimental spectra. The Si(111)-(2$\times$1)
 samples used in the experiments were
cleaved under vacuum and  believed to be single-phase material. However,
scanning tunneling microscopy (STM) and other surface analysis techniques
that are currently available were not available then.

It is clear from Fig.\ \ref{sivbx} that
 the PES calculation recovers the peaks at
$\sim -$1.45 and $\sim -$0.7 eV seen in the experiment.
(The DOS calculation also has features qualitatively
similar to the experiment).
However, the large,
broad intensity seen around $\sim -$3.0 eV,
 is {\it not} reproduced by the PES or the DOS
calculation. This -3 eV broad peak (3eVBP) is less prominent in the
 experimental PES curves reported in subsequent work.\cite{uhrberg}
The bulk spectrum at the ${\bar \Gamma}$-point shows a peak
near -3.5 eV and -5.5 eV. It is very well known
that perfect-clevage surfaces of Si are very
difficult to obtain.
It has been suggested that the 3eVBP is a
result of angle-integrated photoemission from bulk Si, scattered into
the normal direction by surface imperfections.\cite{private} An additional
reason for getting part of the angle-integrated spectrum superimposed on
the angle-resolved spectrum is the effect of diffuse light scattering
in the optical detection system of the particular analyser used in this
specific experiment.\cite{private}

How the angle-integration of the bulk signal by surface imperfections
can lead to a 3eVBP can be understood within the context
of another possible explanation of the 3eVBP that we have explored.
The Si(111)
2$\times$1 structure is metastable with respect to the
7$\times$7 reconstruction. There is no evidence suggesting that
the (2$\times$1) surface reconstructs to the (7$\times$7) form
 at room temperature. However, the
energy dumped on the surface by the 21 eV photon
beam (and the secondary electrons etc., generated during the PES) may
lead to local annealing of the surface structure. Such locally-annealed
 patches may exist in a surface which has already been used
for, say IPES, and then used for obtaining the PES spectrum.
Hence, we have also explored the possibility that the surface may
contain (7$\times$7) patches, now known to be the ground-state of
Si(111).

In order to explore this possibility, we show in Fig.\
\ref{7x7fig}  the experimental angle-resolved
PES from Si(111)-(7$\times$7) for small angles in the range 0-10 deg. We need
to consider only small angles here because the 7$\times$7-phases formed
by relaxation of
the 2$\times$1 phase would be expected to have a set of small orientations
 which averages to zero with reference to the original 2$\times$1-surface.
 Further, the $\bar\Gamma$ point, common to
both (2$\times$1) and (7$\times$7) surfaces, samples only photoelectrons that
are normal to the surface.
The averaged signal from the (7$\times$7) surface for
small angles does show the correct behaviour, giving
a 3eVBP. When the secondary
electron-tail is removed, this broad band is seen to be a small, but
non-negligible,
 component of the whole spectrum. Hence, one possible explanation
of the 3eVBP near $\sim -$3.0 eV would be that (7$\times$7) patches are
present in the cleaved sample believed to be (2$\times$1).

The angle-integration of the bulk signal by surface imperfections
is clearly very analogous to the integration of the angle-resolved
(7$\times$7) data which contain multiple features in the -5 eV to
-2.5 eV range and in the same manner
can lead to a 3eVBP spectral feature. This explanation of the 3eVBP
has the merit of not having to invoke the existence of
7$\times$7-reconstructed patches in the cleaved Si(111) sample.

Nevertheless, al these lead us to emphasizes the importance of having
data from truly well-characterized, clean surfaces when relating theory to
experiment. Further, we note that even if
PES and IPES experiments
were carried out using the same sample, it may still be necessary to
consider local-annealing of the surface
and change of structure when going from
one probe to another.\cite{remark0} Based on the above
discussion and Fig.\ \ref{sicbfig},
we are led to conclude that the sample in IPES -- at least the surface
region probed by the experiment -- was essentially a (2$\times$1) region.

\subsection{PES and IPES for the InP(001)-S surface}
\label{inp-pes}

In contrast to Si, the InP surfaces are much less understood; this is true
in particular of InP(001)-S. Obtaining good STM images or LEED for
 the as-prepared
surface has been difficult. The LEED from the as-prepared surface is
 very poor and the photoemission is almost isotropic, indicative
of poor structural order. In our Raman study of the problem, we recognized
these difficulties and examined the signal arising from the phonons
associated with the splitting of the
{\em underlying} P-layer (just below the In-layer) induced by the surface
reconstruction.\cite{remark}
 Unlike the outermost-S layer, this inner-P layer was expected
to be less influenced by surface contamination. PES and IPES experiments have
been carried out with annealed samples characterized using LEED. However, our
study of the core-level spectra of these samples under annealing showed
that the thermal treatment introduces new complications by favouring
mixed-component phases containing a half-S-half-P-terminated
surface.\cite{tian} Thus, samples annealed at low temperatures
for a short time (e.g,
$\sim$5-10 min.) could easily be an admixture of the InP(001)-S$_2$SS and
InP(001)-(SP)$_2$ phases. The former is the lowest-energy structure for full
S coverage, while the latter is the structure favoured under annealing when
bulk-P atoms migrate to replace the S-atoms which, in turn, diffuse into the
bulk. When higher temperatures are used in annealing, the completely
P-terminated InP(001) phase which is free of sulphur also becomes possible.

Fig.\ \ref{inpsvba}(a) presents our calculation of the PES for both the S$_2$SS
(full line) and the (SP)$_2$ (dashed line) reconstructions of the InP(001)-S
surface. Also included in this figure is the spectrum from bulk InP,
projected onto the surface $\bar\Gamma$ point (dotted line). The experimental
data of Mitchell {\em et al.}\cite{mclean} are displayed as squares in Fig.\
\ref{inpsvba}(b). The PES spectra of each phase (and the bulk) were separately
calculated and aligned to have the same $d$-electron
peak arising from In atoms (which
are not significantly affected by surface reconstructions).

It is clear from Fig.\ \ref{inpsvba} that neither the S$_2$SS spectrum nor the
(SP)$_2$ spectrum, alone, can account for the experimental data, i.e., the
measured spectrum evidently contains features from both surface phases. In
order to show this clearly, we display as the full line in
Fig.\ \ref{inpsvba}(b), a composite spectrum obtained by assuming that
the region of the surface sampled by the light
contains a mixture of S$_2$SS and (SP)$_2$, in proportions of 30\% and
70\% respectively.
 The composite spectrum was positioned (by a rigid shift of the x-axis)
to the
experimental spectrum and scaled to match the intensity of the most prominant
peak. Clearly, a reasonable fit to the experimental data can
be obtained in this way. The experimental spectrum will depend on
details of sample preparation and annealing history. However, an invariant
property would be that a composite theoretical spectrum constructed from the
spectra of the three main annealing components, i.e, S$_2$SS, (SP)$_2$ and
P-terminated InP(001), can always be found to match the essential features of
a given experimental spectrum.

We are therefore led to conclude that the samples examined by Mitchell {\em
et al.}\cite{mclean} using PES
actually contain a mixture of S$_2$SS and (SP)$_2$
phases. This conclusion is in agreement with the observation that the
width of the features in the experimental PES of InP(001)-S is found to be
much broader than from cleaved GaAs or InP(110) surfaces.\cite{private2}

Fig.\ \ref{inpcbfig} presents the results of our IPES calculations
for a sample containing the same mixture of S$_2$SS and (SP)$_2$ phases as we
have just discussed. It is of course not evident that the IPES experiments
were carried out using the same sample that was used for the PES, i.e., with
no further aging or surface modification, and therefore the surface structure
need not be identical. Even with identical samples, the regions of the
surface probed by the photons in PES, and electrons in IPES need not
be identical. The agreement between the calculation and the
experiment, in fact, is very unsatisfactory in the region between 2.5 and 5.5
eV, where the theory predicts a low intensity. However, the fully
P-terminated InP(001) surface is also a possible product of the annealing
process and may be a component of the surface. The PES and STM of the
P-terminated InP(001) has been studied recently.\cite{thesis,wolkow}
 Experiment shows that this phase has
 a prominant  broad absorption band just
in the region 2.5--5.5 eV (dashed curve; Ref.\
\onlinecite{mclean}). This hypothesis has the additional merit
of  providing a natural
explanation for the rather high intensity of the IPES signal in the region
-1.0 to 0.0 eV: there should be very little intensity in this gap region for
the surfaces phases containing no P atoms, as can be seen from the solid
curve. The P-terminated InP(001)-(2$\times$4) surface, in contrast, has a
smaller gap than InP(001)-S, and therefore exhibits a sizable signal in the
-1.0 to 0.0 eV range. This intensity ledge, together with the broad band from
2.5 to 5.0 eV, are therefore consistent with the hypothesis that the sample
contains the InP(001)-P-(2$\times$4)-reconstructed phase as well.

\section{CONCLUSION}

We have analysed the published PES and IPES experimental results for the
S-passivated InP(001) surface using theoretical calculations for various
surface structures which are energetically possible.
 In this analysis we have
borne in mind the fact that the InP(001)-S surface can change its surface
composition and structure under annealing, as shown by core-level
spectroscopy and by total-energy calculations. Our conclusion is that the
experimental results are consistent with a theoretical spectrum obtained from
surfaces containing an admixture of several possible phases which could occur
under thermal annealing. Some of the broad features in the experimental
PES spectrum of Si(111) were discussed in terms of the effect of
surface imperfections or local annealing to form 7$\times$7 patches.
 It is felt
that further theoretical and experimental study of the InP(001)-S surface
would have to await the availability of better characterized surfaces.
However, all the presently-available evidence -- Raman, CLS, PES, IPES
 etc.-- except LEED, seem to be consistent
 with our picture of the as-prepared and
moderately-annealed surfaces being a 2$\times$2-reconstruction, essentially
S$_2$SS for the former and mostly (SP)$_2$ for the latter.


{\em Acknowledgments} --- We thank Franz Himpsel (Wisconsin University)
for useful comments on a first
version of the manuscript, and for several important clarifications.
We also thank Alastair McLean (Queen's University) and Geof Aers (National
Research Council) for many suggestions and comments on the manuscript.
This work was partially supported by grants to LJL from
the Natural Sciences and
Engineering Research Council of Canada and the ``Fonds pour la
formation de chercheurs et l'aide \`a la recherche'' of the Province of
Qu\'ebec.


\begin{figure}
\caption
{
(a) Top view of the fully-relaxed, fully-S-covered InP(001) S$_2$SS surface.
The atom species are identified in the $z$-direction ([001]) layer sequence
shown in (b). The $x$ and $y$ directions are [110] and
[$\bar 1$10], respectively. The atomic positions (\AA) are obtained from
total energy minimization.
\label{Figure1}
}
\end{figure}

\begin{figure}
\caption
{
(a) Top view of the the (SP)$_2$ structure at
 half-sulphur/half-phosphorous coverage. (b) Atomic planes in
the z-direction.
The atomic positions are obtained from
total energy minimization.
\label{Figure2}
}
\end{figure}

\begin{figure}
\caption
{
IPES from the Si(111) surface. The spectrum (PES) calculated using detailed
Kohn-Sham matrix elements is shown as the full line, while the dashed line
gives the result of a DOS calculation.
\label{sicbfig}
}
\end{figure}

\begin{figure}
\caption
{
PES from the Si(111) surface. Experimental results for the (2$\times1$)
structure is
shown  as a solid line.
The theoretical results  for the 16-layer 2$\times1$ slab
(dashed lines)
and for the bulk Si(111) crystal (dot-dashed line) are shown.
\label{sivbx}
}
\end{figure}

\begin{figure}
\caption
{
The solid curve labeled ''7x7-aver'' in the top panel
 is the average over small-angle (0--10 deg.)
 contributions from experimental angle-resolved PES shown in the bottom panel.
\label{7x7fig}
}
\end{figure}

\begin{figure}
\caption
{
Experimental (squares) and theoretical (lines) PES from the InP(001)S surface.
The secondary electron background has been subtracted from the experimental
data. (a) The individual PES spectra from the two surface phases
considered in the text, namely S$_2$SS (full line) and (SP)$_2$ (dashed
line), as well as from bulk InP (dotted line). (b) The full line is a
composite spectrum for a surface containing a mixture
of 30\% S$_2$SS and 70\% (SP)$_2$.
\label{inpsvba}
}
\end{figure}

\begin{figure}
\caption
{
Experimental (squares) and theoretical (full line) IPES from the InP(001)S
surface. Experimental data for the P-terminated InP(001)-(2$\times$4), which
is free of sulphur and produced in higher-temperature anneals, is also shown
(dashed curve).
\label{inpcbfig}
}
\end{figure}


\begin{references}
\bibitem{remark0}
A typical photon flux used in PES experiments would
be of the order of 10$^{11}$ photons falling on
to a 5$\times$5 mm$^2$ region of the
surface. Most of this energy is  carried
away by the photoelectrons.
\end{references}
\end{document}